# Electronically Amplified Electron-Phonon Interaction and Metal-Insulator Transition in Perovskite Nickelates

Yong Zhong[1,2,†,*], Kyuho Lee[1,3,†], Regan Bhatta[4], Yonghun Lee[1,3], Martin Gonzalez[1,3], Jiarui Li[1,3], Ruohan Wang[1,2], Makoto Hashimoto[5], Donghui Lu[5], Sung-Kwan Mo[6], Chunjing Jia[4], Harold Y. Hwang[1,2], and Zhi-Xun Shen[1,2,3,*]

[1]*Stanford Institute for Materials and Energy Sciences, SLAC National Accelerator Laboratory, Menlo Park, CA 94025, USA*

[2]*Departments of Applied Physics, Stanford University, Stanford, CA 94305, USA*

[3]*Departments of Physics, Stanford University, Stanford, CA 94305, USA*

[4]*Departments of Physics, University of Florida, Gainesville, FL 32611, USA*

[5]*Stanford Synchrotron Radiation Lightsource, SLAC National Accelerator Laboratory, Menlo Park, CA 94025, USA*

[6]*Advanced Light Source, Lawrence Berkeley National Laboratory, Berkeley, CA 94720, USA*

[*]Correspondence to: ylzhong@stanford.edu, zxshen@stanford.edu

[†]These authors contributed equally.

The relative role of electron-electron and electron-lattice interactions in driving the metal-insulator transition in perovskite nickelates opens a rare window into the non-trivial interplay of the two important degrees of freedom in solids. The most promising solution is to extract the electronic and lattice contributions during the phase transition by performing high-resolution spectroscopy measurements. Here, we present a three-dimensional electronic structure study of $Nd_{1-x}Sr_xNiO_3$ ($x$ = 0 and 0.175) thin films with unprecedented accuracy, in which the low energy fermiology has a quantitative agreement with model simulations and first-principles calculations. Two characteristic phonons, the octahedral rotational and breathing modes, are illustrated to be coupled with the electron dynamics in the metallic phase, showing a kink structure along the band dispersion, as well as a hump feature in the energy spectrum. Entering the insulating state, the electron-phonon interaction is amplified by strong electron correlations, transforming the mobile large polarons at high temperatures to localized small polarons in the ground state. Moreover, the analysis of quasiparticle residue enables us to establish a transport-spectroscopy correspondence in $Nd_{1-x}Sr_xNiO_3$ thin films. Our findings demonstrate the essential role of electron-lattice interaction enhanced by the electronic correlation to stabilize the insulating phase in the perovskite nickelates.

# Introduction

Metal-insulator transition (MIT) is an important topic in condensed matter physics [1]. The drastic change in resistivity has a direct impact on practical applications. Transition-metal oxides with perovskite structure host many exotic properties associated with the MITs, such as high-$T_C$ superconductivity and colossal magnetoresistance [2,3]. The local $BO_6$ octahedron, as the basic building block of the corner-shared network in the lattice structure, determines most of the electronic and magnetic properties in perovskite oxides. In particular, a variety of ground states and rich phase diagrams emerge from the spin, charge and orbital degrees of the 3$d$ electrons in the B cation [4,5]. Among the material library, perovskite nickelates or rare-earth nickelates, with the formula $R$NiO$_3$ ($R$ = Pr, Nd…, Lu), stand as a canonical example for studying the MIT in transition-metal oxides [6].

The $R$ cation size has a significant impact on the MIT in perovskite nickelates, where the transition temperature can vary by up to hundreds of degrees [7]. The electron configuration of the Ni cation is 3$d^7$, with one electron occupying the degenerate $e_g$ orbitals. Like the high-$T_C$ cuprates, electronic correlations open a charge transfer gap between the Ni-$d$ and O-$p$ bands, giving rise to an antiferromagnetic insulator ground state. Compared to the cuprates, $R$NiO$_3$ has a much smaller charge transfer gap due to the strong $p$-$d$ hybridization. For instance, resistance and optical measurements estimate the gap size of 25 and 200 meV for NdNiO$_3$, respectively, which makes it easy to become a metal at high temperatures [8,9]. Additionally, the small size of the $R$ cation tilts the ideally cubic NiO$_6$ octahedra towards an orthorhombic symmetry, bending the Ni-O-Ni bond angle $\theta$ away from 180°. By considering the hopping parameter $t_{pd}$ between the neighboring Ni and O atoms, the Ni-$d$ orbital bandwidth $W$ is primarily determined via the relation $W \sim t_{pd}^2 cos^2\theta$, which naturally explains the lattice effect on the transition temperature: the smaller the size of the $R$ cation and $W$, the larger the charge transfer gap and the transition temperature [7].

While such basic structural and electronic picture captures the qualitative nature of the MIT in perovskite nickelates, it shows limitations at a deeper level. For example, electron correlation alone cannot explain the resistivity change over five orders of magnitude during the phase transition. Moreover, neutron, x-ray and Raman scattering measurements demonstrate a simultaneous bond disproportionation phenomenon accompanied with the MIT, in which half of the octahedra having a smaller Ni-O bond length and the other octahedra having a larger Ni-O bond length [10-12]. Fig. 1a summarizes the concurrent structural and electronic transitions from an orthorhombic metallic phase to a monoclinic insulating phase in perovskite nickelates. This breathing-type lattice distortion is the key ingredient for establishing microscopic models to interpret the MIT, such as site-selective Mott transition, charge order formation, bi-polaron condensation and structurally triggered mechanism [13-21]. Furthermore, the giant $^{18}$O-$^{16}$O isotope effect on the transition temperature implies the indispensable role of electron-phonon coupling (EPC) to drive the phase transition [22]. Therefore, the inclusion of electron-lattice interaction is necessary to comprehensively understand the MIT of perovskite nickelates.

In addition to the thermal parameter, chemical substitution provides a new dimension to study the transition by tuning the strength of the electron correlation. Both hole- and electron-doped perovskite nickelates have been synthesized by replacing the *R* atom with alkaline earth and lanthanide metals, respectively [23,24]. Comparing with electron dopings, hole dopants suppress the antiferromagnetic insulating state more rapidly. This electron-hole asymmetry is reminiscent of the phase diagram of the cuprates. Here, the lattice size effect is negligible in driving the MIT because the Ni-O-Ni bond angle becomes smaller in electron-doped systems, which in principle should increase the transition temperature. However, experimental results show an opposite trend. In particular, resonant X-ray scattering experiments show that the bond disproportionation is coherently locked to the insulating phase of $Nd_{1-x}Ca_xNiO_3$ thin films [25], again echoing the crucial role of electron-lattice interaction to stabilize the insulating state.

Angle-resolved photoemission spectroscopy (ARPES) has proven to be a powerful tool to study the electronic structure of quantum materials [26]. However, due to the lack of sizeable single crystals and the "uncleavable" property of the perovskite structure, it has been very challenging to perform extensive ARPES measurements on bulk $R$NiO$_3$ samples. Instead, epitaxial thin films have played a vital role to explore the physical properties of $R$NiO$_3$ [27-30]. The transition temperature can be engineered by strain, confinement and interface effects. Although *in situ* ARPES measurements of $R$NiO$_3$ thin films have been reported [31-34], there is no systematic investigation of the three-dimensional electronic structure, especially for the doped counterparts. Hence, a detailed study of the fermiology and electron dynamics is eagerly needed to understand the rich physics of the perovskite nickelates.

In this Article, we report a high-resolution ARPES study of the hole-doped $Nd_{1-x}Sr_xNiO_3$ thin films ($x$ = 0 and 0.175). Three-dimensional electronic structure was precisely constructed by performing photon-energy dependent measurements. The Fermi surface topology and low energy band dispersions are quantitatively described within the framework of a two-band tight binding model, which is further confirmed by first-principles calculations. Both rotational and breathing phonon modes of the NiO$_6$ octahedra are identified in the metallic phase, showing a kink structure along the band dispersion and a polaron feature in the energy spectrum. Moreover, by tracking the spectral evolution across the transition, we observe a salient spectral weight transfer accompanied by a gap formation in the insulating state. This strong electron correlation substantially reinforces the electron-lattice interaction to freeze the polarons in the ground state.

## Methods

Nd$_{1-x}$Sr$_x$NiO$_3$ ($x$ = 0, 0.175) epitaxial thin films were synthesized by the pulse-laser deposition method on (LaAlO$_3$)$_{0.3}$(Sr$_2$TaAlO$_6$)$_{0.7}$ (001) substrate [35]. The film thickness is about 6 nm. After the growth, the samples were transferred to a commercial oxide molecular beam epitaxy system (Veeco GEN930), which is *in situ* connected to the ARPES end station at Stanford Synchrotron Radiation Lightsource beamline 5-2 equipped with a Scienta DA30-L analyzer. We annealed the thin films at 500 °C for 20 min under a base pressure of 1 × 10$^{-5}$ Torr purified ozone to ensure a clean surface for ARPES study [36]. The base pressure during the ARPES measurements was better than 3 × 10$^{-11}$ Torr. Linear-horizontally polarized light within a photon energy range 60 – 180 eV was used to explore the three-dimensional electronic structure. The measurement temperature was 10 K on $x$ = 0.175 sample, and 80 – 180 K on $x$ = 0 sample.

First-principles calculations for electronic structures were performed in the framework of DFT using Projected Augmented Wave (PAW) pseudopotentials [37] as implemented within the Vienna Ab initio Simulation Package (VASP) [38]. The calculations employed the Generalized Gradient Approximation (GGA) using the PBE exchange-correlation functional [39], as well as the DFT+U approach using the spherically averaged form of the rotationally invariant effective U parameter [40]. The electronic structure calculation as shown in the main text is implemented at A-type AFM structure with U = 4.0 eV on the correlated Ni 3$d$ orbitals. 24 × 24 × 12 Monkhorst-Pack grids centered at Gamma point is used in the self-consistent calculation [41]. We implemented the method of Methfessel-Paxton of order 4 for smearing, with sigma = 0.2eV. For the cubic structure calculation as shown in the supplementary information, we chose 24 × 24 × 24 Monkhorst-Pack grid centered at Gamma point and U = 1.5 eV for the DFT+U calculation [18].

Fig. 1b displays the resistivity measurements on Nd$_{1-x}$Sr$_x$NiO$_3$ thin films. The resistivity curve of $x$ = 0 sample shows a thermally driven MIT with a hysteretic loop, confirming a first-order phase transition. Chemical substitution of Nd with Sr suppresses the insulating ground state, providing a new knob to control the transition at low temperature regime. Considering the surface sensitivity of ARPES measurements, we used reflection high-energy electron diffraction (RHEED) to monitor the surface quality during the ozone-annealing process (Fig. 1c). The 2 × 2 surface reconstruction in the undoped sample arises from the catastrophe effect of a polarized termination [42,43]. Hole dopants neutralize this charge imbalance, recovering the 1 × 1 RHEED pattern for the $x = 0.175$ sample. After ozone annealing, all samples have sharp RHHED streaks, which are very important to obtain high-quality data in the following ARPES measurements.

## Results

**Three-dimensional Fermi surface**

Fig. 2a displays a schematic of the three-dimensional Fermi surface of the nickelates in the reference cubic structure, consisting of an electron pocket centered at the Γ point and a hole pocket centered at the A point. Theoretical studies demonstrate that the two Ni $e_g$ orbitals contribute dominantly to the low energy physics in the system [15, 44]. Experimentally, the full set of ($k_x$, $k_y$, $k_z$, $E$) information is crucial to build the three-dimensional electronic structure. In ARPES measurements, the in-plane momentum $\mathbf{k}_{\parallel}$ and out-of-plane momentum $k_z$ can be obtained by the following relations

$$|\mathbf{k}_{\parallel}| = \sqrt{2m(h\nu - \phi - E_B)}\sin\theta/\hbar \tag{1}$$

$$k_z = \sqrt{2m[(h\nu - \phi - E_B)\cos^2\theta + V_0]}/\hbar \tag{2}$$

where $m$ is the free electron mass, $h\nu$ is the incident photon energy, $\phi$ is the analyzer work function, $E_B$ is the binding energy, $\theta$ is the polar angle with respect to the surface normal, and $V_0$ is the inner potential which can be extracted by analyzing the experimental data and considering the periodicity of the Brillouin zone (BZ). By varying the incident photon energy, we can cut different slices of the three-dimensional Fermi surface (Fig. 2b). The variation of the $k_z$ value in the neighboring BZs can be as large as 0.2 $\pi/a$, where $a$ is the in-plane lattice constant. Figs. 2c and 2d show the two-dimensional projected Fermi surfaces at $h\nu$ = 80 eV and 155 eV, respectively. For simplicity, we label the high symmetry points as Γ', X' and M' to distinguish them from the points in the $k_z = 0$ plane. Benefitted from the high-resolution spectroscopy, now we can obtain many new aspects of the Fermi surface structure that have not been reported in previous studies [31-34]. For example, we can clearly map the electron pocket around the Γ point, as well as the asymmetric Fermi surface at higher-order BZs that arise from the variation of $k_z$ value along the $\mathbf{k}_{\parallel}$ direction.

To parametrize the Fermi surface, we consider the following tight binding model [44],

$$H_{tb} = -\sum_{ij} t_{ij}^{ab} c_{ia}^+ c_{jb} - \mu \tag{3}$$

Where $i, j$ are site indices, $a, b = 1, 2$ are orbital indices denoting the two $e_g$ states with $d_{x^2-y^2}$ and $d_{z^2}$ orbitals, and $\mu$ is the chemical potential. Only nearest ($t_{\pm ax}$, $t_{\pm ay}$, $t_{\pm az}$) and next-nearest neighbor ($t_{\pm ax \pm ay}$, $t_{\pm ax \pm az}$, $t_{\pm ay \pm az}$) hopping processes are considered in our model,

$$t_{\pm ax} = \frac{t_1}{4}\begin{pmatrix} 1 & -\sqrt{3} \\ -\sqrt{3} & 3 \end{pmatrix}, t_{\pm ay} = \frac{t_1}{4}\begin{pmatrix} 1 & \sqrt{3} \\ \sqrt{3} & 3 \end{pmatrix}, t_{\pm az} = t_1\begin{pmatrix} 1 & 0 \\ 0 & 0 \end{pmatrix} \tag{4}$$

$$t_{\pm ax \pm ay} = \frac{t_2}{2}\begin{pmatrix} 1 & 0 \\ 0 & -3 \end{pmatrix}, t_{\pm ax \pm az} = \frac{t_2}{2}\begin{pmatrix} -2 & \sqrt{3} \\ \sqrt{3} & 0 \end{pmatrix}, t_{\pm ay \pm az} = \frac{t_2}{2}\begin{pmatrix} -2 & -\sqrt{3} \\ -\sqrt{3} & 0 \end{pmatrix} \quad (5)$$

where $t_1$ and $t_2$ are the hopping parameters. The basis [1, 0] and [0, 1] are the $d_{z^2}$ and $d_{x^2-y^2}$ orbitals. We compute the ARPES intensity $I(E, k_x, k_y)$ by averaging over $k_z$ using a Lorentzian convolution with a full width at half-maximum (FWHM) given by $\Delta k_z = 1/\lambda$,

$$I(E, k_x, k_y) = \int_{k_z^0 - \Delta k_z/2}^{k_z^0 + \Delta k_z/2} f(E) \frac{A(E,k)}{(k_z - k_z^0)^2 + (\Delta k_z/2)^2} \quad (6)$$

Here, $f(E)$ is the Fermi-Dirac function, $k_z^0$ is the nominal $k_z$ value probed at the selected photon energy, $\lambda$ is the photoelectron mean free path, and $A(E, \mathbf{k})$ is the spectral function

$$A(E, \mathbf{k}) = \sum_i -\frac{1}{\pi} \frac{\Sigma''(E)}{(E - \varepsilon(\mathbf{k}) - \Sigma'(E))^2 + (\Sigma''(E))^2} \quad (7)$$

where $\varepsilon(\mathbf{k})$ is the electron band energy at momentum $\mathbf{k}$ and $\Sigma'(E)$ and $\Sigma''(E)$ are the real and imaginary parts of the self-energy, respectively. All the parameters of the model simulations are listed in Table 1.

The simulated Fermi surfaces with $h\nu$ = 80 eV and 155 eV are displayed in Figs. 2e and 2f. To better benchmark with the theory, we employ the momentum-distribution-curve (MDC) analysis to extract the experimental Fermi momenta $k_F$s, as labelled by the red circles in Fig. 2. There is a quantitative consistency between the model simulations and experimental observations. Additionally, we present more Fermi surface simulations at $h\nu$ = 70 eV, 75 eV, 90 eV, 150 eV and 160 eV. All of them agree well with the experimental results (Fig. S1). Fig. 2g and 2h show the energy-momentum (*E-k*) image plot and the corresponding energy-distribution-curve (EDC) of the valence band. The sharp peak near the Fermi level is mostly contributed by the Ni-$e_g$ states, while the shoulder feature around ~ 0.8 eV and the broad peak with onset at ~ 2.3 eV arise from the Ni-$t_{2g}$ and the O-*2p* states, respectively [45]. Similar spectral features were also observed at $h\nu$ = 80 eV (Fig. S2), indicating that our spectroscopy measurements elucidate the intrinsic valence band information of the perovskite nickelates.

In addition to the *x* = 0.175 sample, we also explored the Fermi surface contours of the parent compound NdNiO$_3$, as shown in Fig. S3. The basic topology is nearly the same for both samples: consisting of a small electron pocket at the BZ center and a large hole pocket around the BZ corner. However, the sizes of the pockets are slightly different in these two samples. According to the Luttinger theorem, we can calculate the doping levels by counting the filled states enclosed by the Fermi surface pockets. Table 2 summarizes the key results: the calculated doping levels $x_L$ = 0.006 and 0.172 are consistent with the nominal values of *x* = 0 and 0.175. Moreover, we find that the hole pocket can be effectively changed by the Sr dopants, while the electron pocket is immune to the carrier doping. Thermodynamic measurements show negative Seebeck coefficients in NdNiO$_3$ and PrNiO$_3$ compounds, demonstrating the nature of electron-type charge transport in the metallic phases [8,46]. Specifically, our ARPES measurements obtain an effective electron concentration of 0.05-0.06 $e$/Ni atom in $Nd_{1-x}Sr_xNiO_3$ thin films, implying the bad metal property of the perovskite nickelates [47].

**Low energy electronic structure**

The intrinsic three-dimensionality of the perovskite structure significantly affects the band dispersions at different $k_z$ values. For example, Fig. 3a displays the experimental $E$-$k$ image plot along the BZ boundary at $hv = 80$ eV. Although the variation of $k_z$ substantially changes the bandwidths in the first and second BZs, the tight binding simulation impressively reproduces the band structures with quantitative consistency (Fig. 3b), implying the feasibility of the two-band model to capture the low energy physics of perovskite nickelates. Moreover, the high-resolution spectroscopy cuts at $hv = 150$ eV allow us to obtain the degenerate $e_g$ bands simultaneously along the diagonal direction of the BZ, as displayed in Fig. 3c. Again, most of the essential features are well described by the model simulations (Fig. 3e). For example, the $d_{z^2}$ orbital contributes predominantly to the hole pocket, while the electron pocket comes from the hybridization of the two $e_g$ orbitals. The residual intensity in the high energy regime (also called the waterfall feature) is suggested to come from the many-body effects, as extensively discussed in the high-$T_C$ cuprates and other correlated oxides, although no consensus has yet been reached [48,49].

We can further analyze the orbital-dependent renormalization effects by comparing with the density functional theory (DFT) calculations. Fig. 3f displays the DFT results along the same cuts and $k_z$ value. Although qualitative agreement is achieved, the bandwidths in DFT calculations are obviously larger than both the experimental measurements and tight binding simulations. We find that the measured electron-type and hole-type carriers are renormalized by factors of about 2.2 and 4.6 to match the calculated band dispersions. The results are consistent with previous study on $LaNiO_3$, where the renormalized value is about 3 [45]. More details of the orbital-dependent renormalizations are summarized in Table 3. It should be noticed that we did not observe the non-dispersive band along Γ' to X' direction, which has been predicted by both theoretical methods. There are two possible reasons for this discrepancy. First, the tight binding model assumes a cubic symmetry rather than the realistic orthorhombic symmetry. The lowered lattice symmetry in real materials can push this non-dispersive band away from the Fermi level. Indeed, the DFT calculations on a cubic structure show degenerate non-dispersive bands from Γ to X (Fig. S5), agreeing well with the model simulations. Second, the non-dispersive band in DFT calculations is mostly contributed by the O-$2p$ orbital, which is not sensitive to the experimental photon energies $hv = 60 – 180$ eV.

A direct investigation of the coupling interactions between electrons and collective excitations provides important information beyond the theoretical predictions. The left panel of Fig. 3d displays the MDC analysis of the band dispersion for X' to M' cut: a kink around $E_B$ = 36 meV is unambiguously detected. The kink structure originates from the electron-boson coupling, such as magnetic order or phonon [50,51]. We can rule out the magnetic order scenario because the metallic phase is in a paramagnetic state. Then, a likely candidate is the phonon mode of the $NiO_6$ octahedra. Raman studies showed that the rotational octahedra have a characteristic phonon energy of 36 meV [12,52], in good agreement with the kink energy in our measurements. In the electron-phonon model, the EPC strength $\lambda$ is related to the ratio of the Fermi velocity $v_F$ and the bare-band velocity $v_B$: $\lambda = v_B/v_F - 1$. Here, we can estimate $\lambda = 0.6$ in $Nd_{1-x}Sr_xNiO_3$ (Fig. 3d), implying a weak EPC in the metallic phase. Furthermore, we can extract another phonon mode through the self-energy analysis. The real and imaginary parts of self-energy are shown in the right panel of Fig. 3d: in addition to the primary peak at 36 meV, a second peak around 60 meV is clearly illustrated for the imaginary part, which is also confirmed as a step-function feature in the real part. This phonon mode corresponds to the breathing distortion of the $NiO_6$ octahedra [12,52], as will be discussed in detail below.

## Discussions

The signature of large polaron is clearly detected in the metallic phase. Fig. 4b displays the small-scale energy spectra of the $x$ = 0.175 sample: on top of the pronounced quasiparticle peak, the spectra show a hump feature around 85 meV that is attributed to the EPC. The peak-hump interval is the related phonon energy $\hbar\omega_0$ = 65 meV, consistent with the breathing mode and the self-energy analysis (Fig. 3d). To better understand the phase transition in the perovskite nickelates, Fig. 4a displays the temperature-dependent large-scale energy spectra in $NdNiO_3$ (more data in Fig. S6). The metallic spectrum at high temperature resembles that of the doped sample (Fig. 2h), demonstrating the same microscopic mechanism for temperature-driven and doping-driven MIT's in the perovskite nickelates. Compared with the metallic state, the insulating spectrum exhibits a gap feature and a salient spectral weight transfer from the $e_g$ and $t_{2g}$ orbitals to the deep O-$2p$ states, as depicted by the arrow in Fig. 4a. Here, we use the shift of the $t_{2g}$ shoulder in the metallic and insulating states to estimate the insulating gap: the gap size is about 300 meV, comparable with the results from optical measurements [9]. In particular, the spectra evolution is reminiscent of the behavior in magnetoresistive manganites and high-$T_C$ cuprates [53,54], implying the formation of small polarons in the insulating state [55,56]. Therefore, we reveal the MIT triggered by the transition of large polarons to small polarons from a spectral perspective: the well-defined quasiparticles and hump structure in the metallic state, as well as the broad line shapes and missing quasiparticles in the insulating phase.

The EPC in relation to the MIT is further demonstrated by tracking the temperature-dependent $E$-$k$ image plot and small-scale energy spectra in NdNiO$_3$ (Fig. 4c and 4d). First, we reproduced the resistivity measurements by analyzing the quasiparticle residue $Z$, which is defined as the spectral integration from $E_F$ to the kink energy (shaded region in Fig. 4d). According to the empirical "$\sigma_{dc} \propto Z$" law [53], the resistivity curve with the landmark of hysteresis loop is successfully repeated, as shown in Fig. 4(e). This transport-spectroscopy correspondence establishes the reliability of spectral characterization of the MIT in the perovskite nickelates. Second, the large polaron feature manifests in the metallic state and disappears in the insulating phase (more information in Fig. S7). Since the MIT is coupled to a change of lattice symmetry, the enhanced lattice distortion traps the mobile polarons and transforms them into small polarons. In fact, the strength of electron-lattice interaction can be effectively amplified by the strong electron correlation, as supported by the DMFT calculations in high-$T_C$ cuprates [56].

A unified picture of the underlying mechanism for the MIT can be summarized in Fig. 4f. The phase transition is coherently locked to a structural change, a common feature of many MIT systems. For example, VO$_2$ has a rutile to monoclinic structural transition [57], and Ta$_2$NiSe$_5$ accompanies with an orthorhombic to monoclinic change [58]. In the metallic state, the weak EPC slightly alters the motion of electrons and forms large polarons. As entering the insulating phase, the reduced lattice symmetry can enhance the EPC strength, freezing the mobile polarons into localized ones around the transition metal cations. Such a large to small polaron transition has been reported in perovskite manganites [59], which shows a similar lattice effect on the MIT temperature [60]. For the perovskite nickelates, both thermal and doping parameters can tune the EPC strength, which in turn modulates the transition temperature. In our photoemission studies, we found signatures of both rotational and breathing phonon modes associated with the NiO$_6$ octahedra, indicating their influence on the electronic behavior of the perovskite nickelates. DFT calculations demonstrate that these two phonons work together to drive the MIT: the rotational distortion plays a key role to stabilize the softening of the breathing mode, which is the main driving force for the phase transition [18]. Lastly, we want to emphasize the multi-channel feature of our picture: the electron-lattice interaction can reshape the local electronic correlation by tuning the octahedral rotation motion, whereas the strong electronic correlation can reinforce the EPC.

## Conclusion

In summary, we established the three-dimensional electronic structure of the perovskite nickelates. The two-band model quantitatively captures the low energy physics and electron dynamics. The synergy of electron-lattice and electron-electron interactions drives the MIT, accompanied by the transition from large polarons to small polarons. Furthermore, our simple and reliable method for obtaining clean surfaces broadens the scope of spectroscopic studies of transition metal oxides.

## Acknowledgements

The authors are grateful to Tom Devereaux, Brian Moritz, Yijun Yu, Yujun Deng and Matthias Hepting for fruitful discussions. This study at the SSRL/SLAC is supported by the U.S. Department of Energy, Office of Science, Office of Basic Energy Sciences under contract no. DE-AC02- 76SF00515.

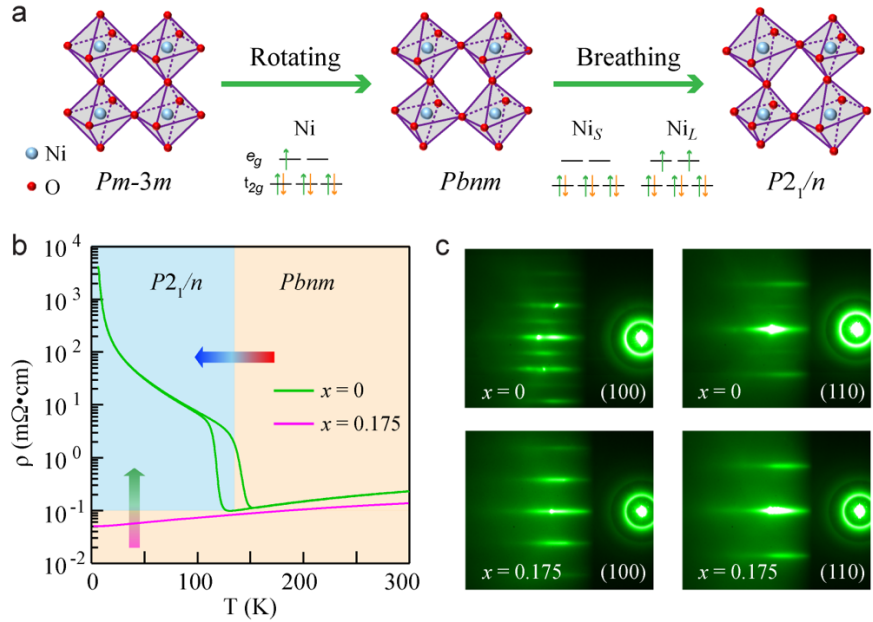

FIG. 1 Metal-insulator transition (MIT) in the perovskite nickelates. (a) Evolution of the crystal structures in the perovskite nickelates. The standard perovskite structure should have a cubic symmetry ($Pm$-$3m$). Due to the small size of rare-earth atom, the NiO$_6$ octahedra tilt to the orthorhombic symmetry ($Pbnm$) in the metallic state. The phase transition accompanies with a bond disproportionation phenomenon, in which the crystal structure lowers to a monoclinic symmetry ($P2_1/n$) in the insulating state. The inset panel describes the electron configurations of Ni cations in the evolution of the crystal structures. (b) Resistivity measurements of Nd$_{1-x}$Sr$_x$NiO$_3$ ($x$ = 0, 0.175) thin films. The orange area represents the orthorhombic symmetry in the metallic phase, while the blue area denotes the monoclinic symmetry in the insulating state. Both thermal fluctuation (horizontal arrow) and chemical doping (vertical arrow) can drive the MITs in Nd$_{1-x}$Sr$_x$NiO$_3$. (c) RHEED patterns of $x$ = 0, 0.175 samples after the ozone annealing process. The 2 × 2 reconstruction in undoped sample arises from the catastrophe effect with a polarized surface termination.

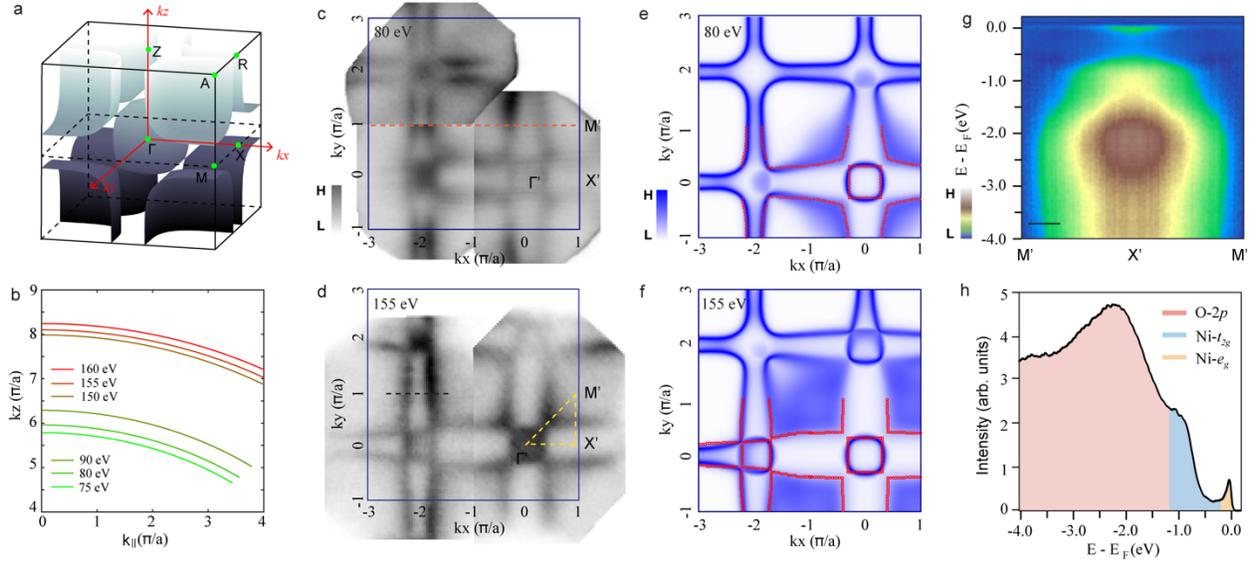

FIG. 2 Three-dimensional Fermi surface of the perovskite nickelates. (a) Schematic diagram of the Fermi surface: two enclosed pockets centered at the Γ and A points, respectively. (b) $k_z$ - $k_{||}$ relations at different photon energies $h\nu$ = 75 eV, 80 eV, 90 eV, 150 eV, 155 eV and 160 eV. (c, d) The experimental Fermi surfaces at $h\nu$ = 80 eV and 155 eV. The integration window is 20 meV around $E_F$. (e, f) The simulated Fermi surfaces at $h\nu$ = 80 eV and 155 eV. The red circles are experimental $k_F$ determined by the MDC analysis. (g) $E$-$k$ image plot of the valence band along the second BZ boundary (black dashed line in Fig. 2d). (h) The corresponding EDC spectrum of the valence band. The quasiparticle feature near $E_F$ is mostly contributed by the Ni-$e_g$ orbitals. The peaks around 0.8 eV and 2.3 eV come from the Ni-$t_{2g}$ and the O-$2p$ orbitals, respectively. All the measurements were performed on the $x$ = 0.175 sample at 10 K.

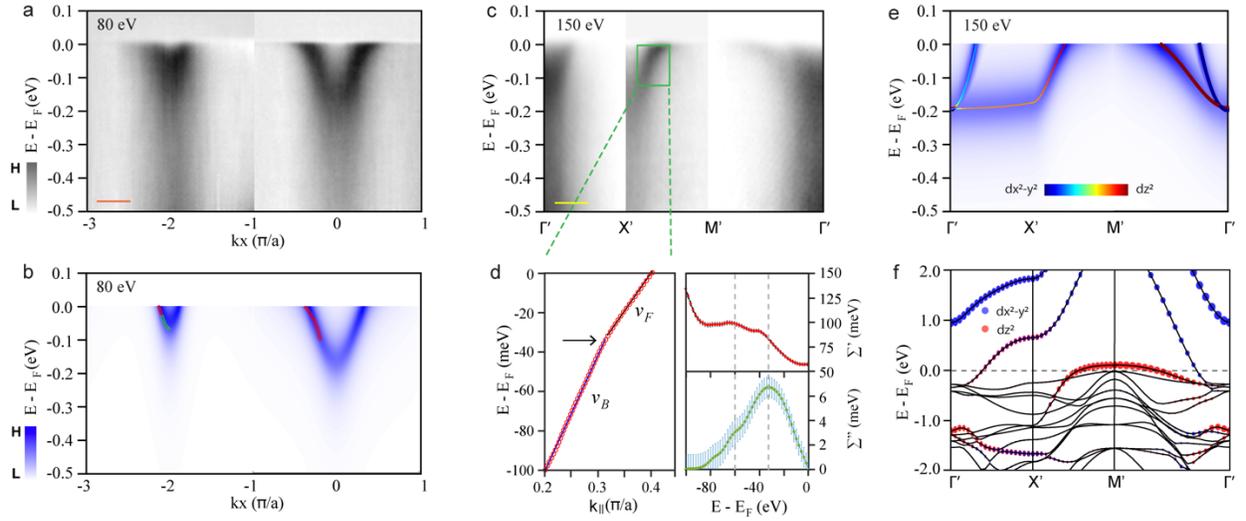

FIG. 3 Low-energy electronic structure of the perovskite nickelates. The momentum directions are labelled by the same-color lines shown in Fig. 2. (a, b) The experimental and simulated *E-k* image plots along the BZ boundary (orange dashed line in Fig. 2c) at $h\nu$ = 80 eV. The red and green circles are the experimental momenta determined by the MDC and EDC analysis, respectively. (c) Experimental *E-k* image plot along the high-symmetry cuts (yellow dashed line in Fig. 2d) at $h\nu$ = 150 eV. (d) Left panel: enlarged band dispersion of the green box shown in Fig. 3c, in which a kink structure around 36 meV is clearly observed. The red circles are experimental $k_\parallel$ determined by the MDC analysis. Fermi velocity $v_F$ and bare-band velocity $v_B$ are defined as the black and blue lines, respectively. Right panel: the energy-dependent real (top) and imaginary (bottom) parts of the self-energy: two phonon modes at 36 meV and 60 meV are illustrated. (e) The corresponding *E-k* image simulation along the same high-symmetry cuts in Fig. 3c. The relative components of $d_{z^2}$ and $d_{x^2-y^2}$ orbitals are labeled by the color scale. (f) DFT calculations of the band structure along the same high-symmetry cuts in Fig. 3c. There is a remarkable agreement between the experimental observations and the DFT results. All the measurements were performed on the *x* = 0.175 sample at 10 K.

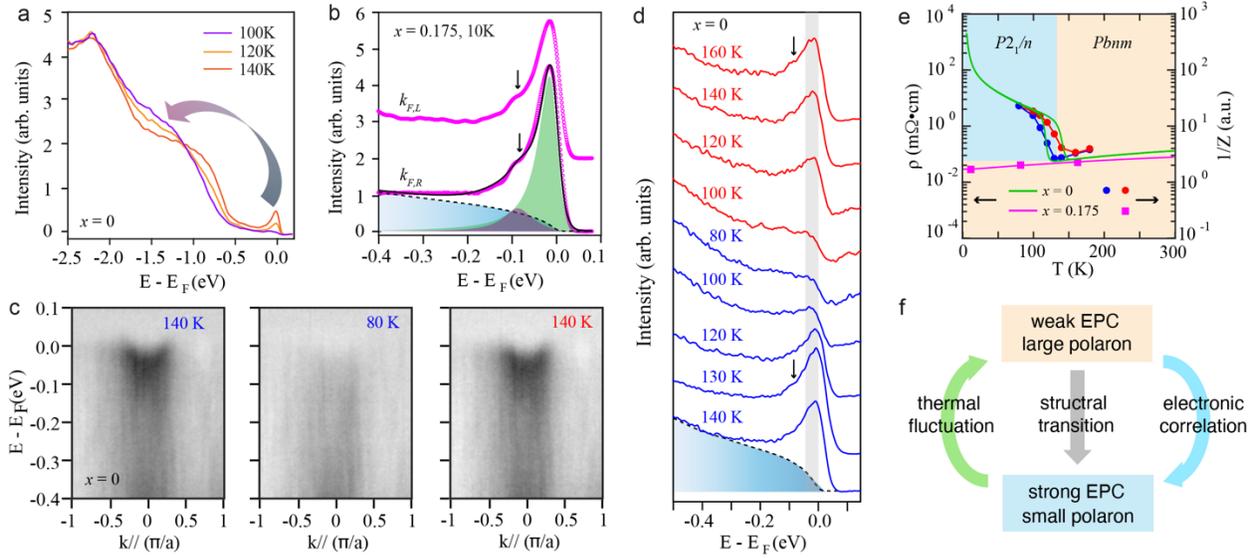

FIG. 4 The role of electron-phonon coupling (EPC) on the MIT. (a) Temperature-dependent EDC spectra of the valence band across the MIT on the $x = 0$ sample. The spectral line shape in the metallic phase (T=140 K) is nearly the same as that of the $x = 0.175$ sample (Fig. 2h), implying the same microscopic mechanism of the thermal-driven and doping-driven MITs in the perovskite nickelates. In the insulating phase, the spectral intensity of the quasiparticle feature near $E_F$, as well as the Ni-$t_{2g}$ peak around 0.8 eV, are drastically suppressed and then transferred to the deeper O-$2p$ states, which is reminiscent of the spectral weight redistribution in the underdoped regime of high-$T_C$ cuprate superconductors. (b) Polaronic signature in the metallic state of the $x = 0.175$ sample. In addition to the strong quasiparticle peak near $E_F$, there is an obvious hump feature (denoted by the black arrows) in the EDC curves taken at $k_F$, which can be seen as a fingerprint of large polarons in the metallic phase. (c) $E$-$k$ image plots across the MIT on the $x = 0$ sample. (d) Temperature-dependent EDC spectra collected at $k_F$ for $x = 0$ sample. The faint hump feature (denoted by the black arrows) can still be observed in the metallic state. Quasiparticle residue Z is extracted by subtracting a Shirley-type background (blue shaded area) and integrating the spectral weight intensity in the energy window of [0, -36 mV] (gray shaded area). The blue color represents the cooling process, while the red color denotes the heating process. (e) Comparison of the resistivity curves and the 1/Z values on the $x = 0$ and 0.175 samples. There is a good consistency between the transport measurement and the spectroscopy analysis. (f) Schematic of the microscopic mechanism of the MIT. In the metallic phase, the EPC is modest to form large polarons, manifesting as the hump feature (coupling to the octahedral breathing mode) and the kink structure (coupling to the octahedral rotation mode). Then, the EPC becomes stronger in the insulating state because of the structural change accompanying with the MIT. The lower crystal symmetry can enhance the EPC to trap the electrons locally. Moreover, the strong correlation will amplify the EPC to support the formation of polaron solid phase in the ground state.

TABLE 1 Tight-binding parameters

| $t_1$ | $t_2$ | $\mu$ for x =0.175 sample | $\mu$ for x = 0 sample | $\lambda$ | $V_0$ |
|---|---|---|---|---|---|
| 0.18 eV | 0.1 $t_1$ | 3.7 $t_1$ | 3.44 $t_1$ | 5 Å | 12 eV |

TABLE 2 Luttinger theorem analysis for the Fermi surface pockets

| Sample | Hole doping | electron doping | ARPES doping |
|---|---|---|---|
| x =0.175 | 21.74% | 4.48% | 17.26% |
| x = 0 | 6.71% | 6.15% | 0.56% |

TABLE 3 Renormalization effects for different orbitals

| k cut | Orbital | Experimental bandwidth | DFT bandwidth | Renormalization |
|---|---|---|---|---|
| $\Gamma' - X'$ | $d_{z^2}$ | 0.15 eV | 0.33 eV | 2.2 |
| $\Gamma' - M'$ | $d_{x^2-y^2}$ | 0.15eV | 0.33 eV | 2.2 |
| $\Gamma' - M'$ | $d_{z^2}$ | 0.15 eV | 0.33 eV | 2.2 |
| $X' - M'$ | $d_{z^2}$ | 0.22 eV | 1 eV | 4.6 |

# Supplementary Materials

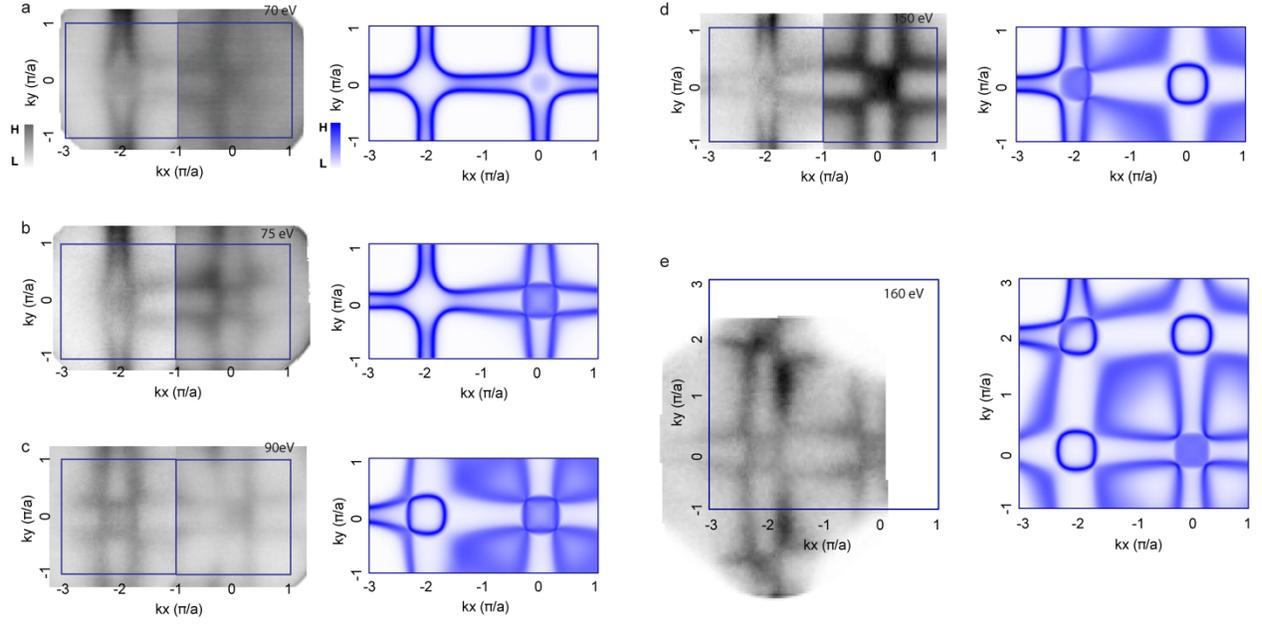

FIG. S1 (a-e) The Experimental and simulated Fermi surface contours at $h\nu$ = 70 eV, 75 eV, 90 eV, 150 eV and 160 eV. All the measurements were performed on $x$ = 0.175 sample. The experimental temperature was 10 K. All the parameters we used in the tight binding simulations are listed in Table 1.

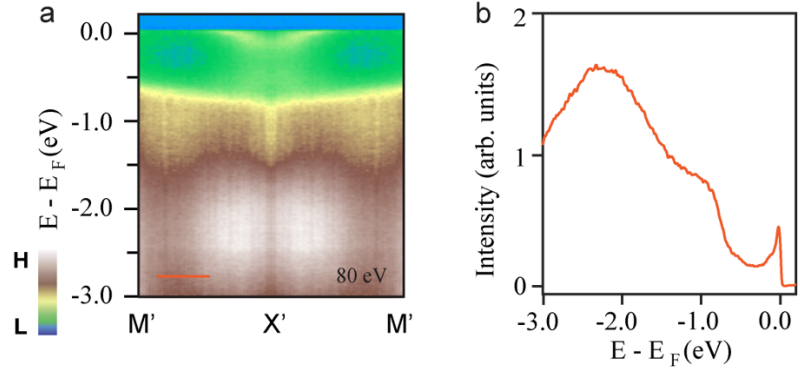

FIG. S2 (a) *E-k* image plot of the valence band along the first BZ boundary (labeled by orange dashed line in Fig. 2c) at $h\nu = 80$ eV. (b) The corresponding EDC spectrum of the valence band is the same as the one in Fig. 2h.

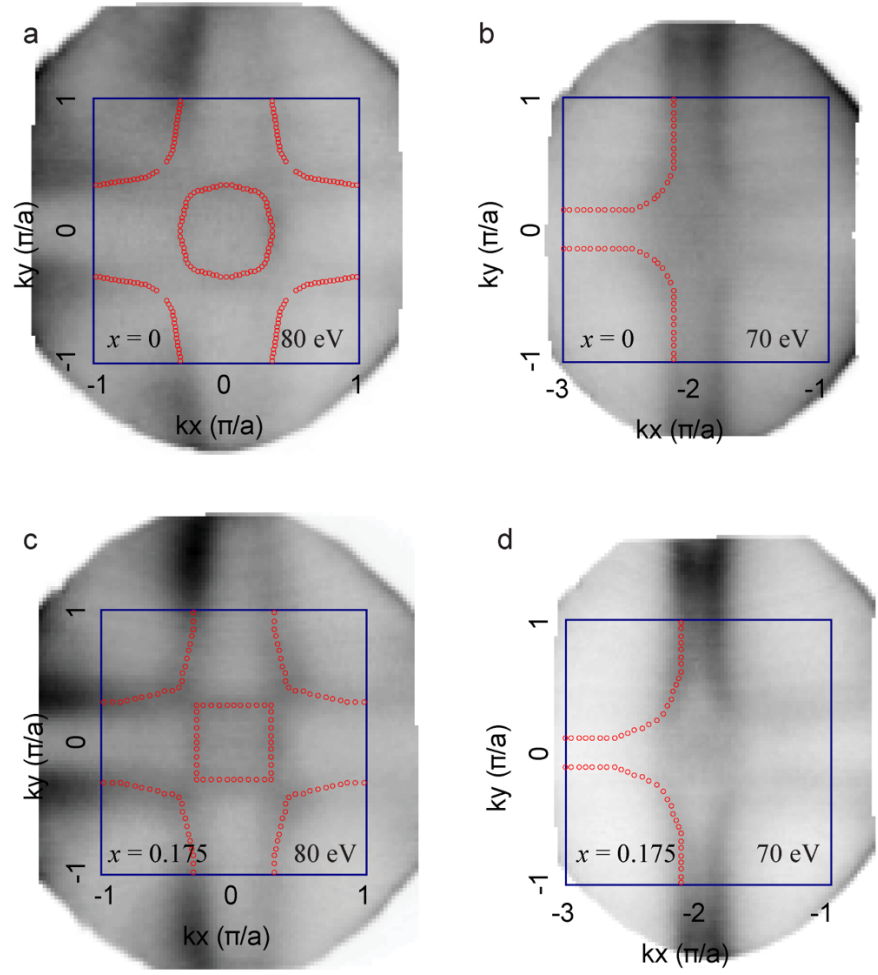

FIG. S3 (a, b) Fermi surface contours of $x = 0$ sample in the first and second BZs. (c, d) Fermi surface contours of $x = 0.175$ sample in the first and second BZs. The red circles are experimental $k_F$ determined by MDC analysis.

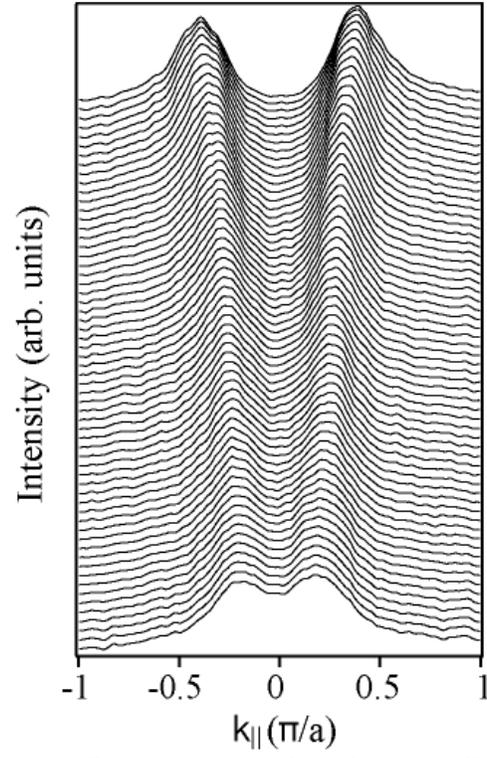
FIG. S4 The corresponding MDC curves along the X'-M' direction in Fig. 3d.

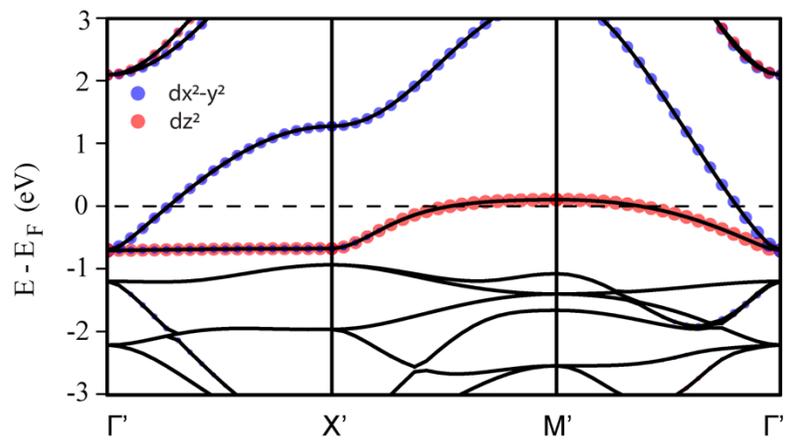

FIG. S5 DFT calculations in the cubic structure.

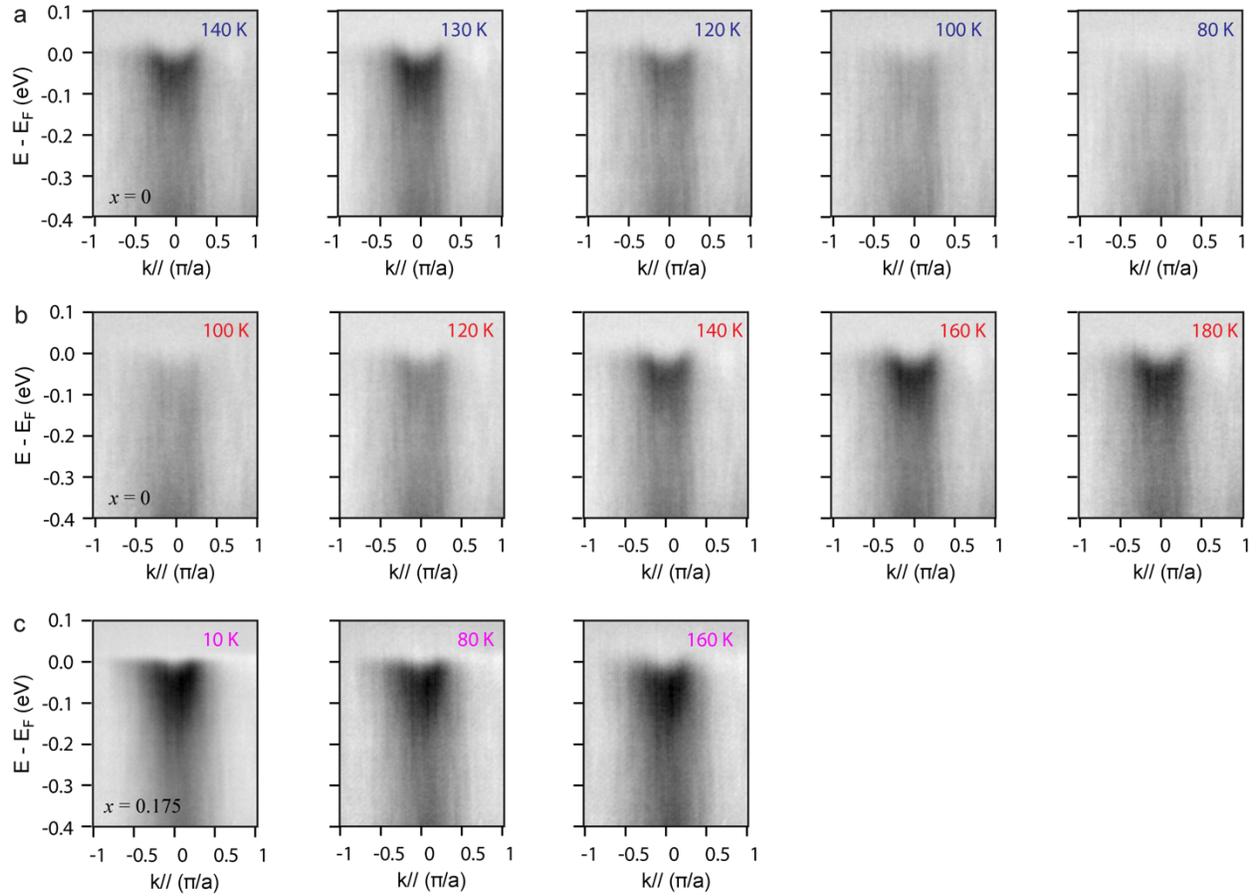

FIG. S6 (a) Temperature-dependent *E-k* image plots of *x* = 0 sample in the cooling process. (b) Temperature-dependent *E-k* image plots of *x* = 0 sample during the heating process. (c) Temperature-dependent *E-k* image plots of *x* = 0.175 sample.

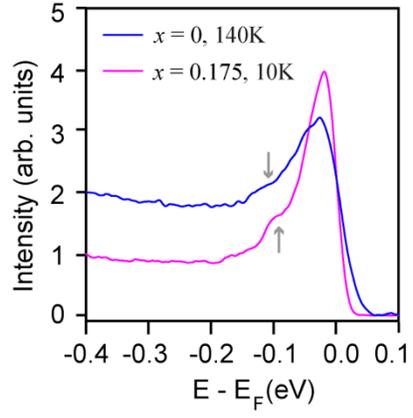

FIG. S7 The metallic EDC spectra in $x = 0$ and $x = 0.175$ samples. The hump feature (gray arrows) is observed for both cases, indicating the formation of large polarons in the temperature-driven and doping-driven metallic states.